\documentstyle[preprint,prc,aps,eqsecnum,epsf]{revtex}
\tighten
\begin{document}
\draft

\title{Parity-Violating Interaction Effects I: the
Longitudinal Asymmetry in $p$$p$ Elastic Scattering}
\author{J.\ Carlson}
\address{Theoretical Division, Los Alamos National Laboratory, Los Alamos, New Mexico 87545}
\author{R.\ Schiavilla}
\address{Jefferson Lab, Newport News, Virginia 23606 \\
         and \\
         Department of Physics, Old Dominion University, Norfolk, Virginia 23529}
\author{V.R.\ Brown}
\address{
         Department of Physics, Massachusetts Institute of Technology,
         Cambridge, MA 02139 \\
         and \\
         Department of Physics, University of Maryland, College Park, MD 
         20742 
         }
\author{B.F.\ Gibson}
\address{Theoretical Division, Los Alamos National Laboratory, Los Alamos, New Mexico 87545}
\date{\today}
\maketitle

\begin{abstract}
The proton-proton parity-violating longitudinal
asymmetry is calculated in the lab-energy range 0--350 MeV, using
a number of different, latest-generation strong-interaction
potentials--Argonne $v_{18}$, Bonn-2000, and Nijmegen-I--in combination
with a weak-interaction potential consisting of $\rho$- and $\omega$-meson
exchanges--the model known as DDH.  The complete scattering problem
in the presence of parity-conserving, including Coulomb,
and parity-violating potentials is solved in both configuration- and
momentum-space.  The predicted parity-violating asymmetries are found
to be only weakly dependent upon the input strong-interaction potential adopted
in the calculation.  Values for the $\rho$- and $\omega$-meson
weak coupling constants $h^{pp}_\rho$ and $h^{pp}_\omega$
are determined by reproducing the measured asymmetries at 13.6 MeV,
45 MeV, and 221 MeV.
\end{abstract}

\section{Introduction}
\label{sec:intro}

A new generation of experiments have recently
been completed, or are presently under
way or in their planning phase to study the effects
of parity-violating (PV) interactions in $p$$p$
elastic scattering~\cite{Berdoz01}, $n$$p$ radiative
capture~\cite{Snow00} and deuteron electro-disintegration~\cite{Jlab01}
at low energies.  There is also considerable interest
in determining the extent to which PV interactions can
affect the longitudinal asymmetry measured by the
SAMPLE collaboration in quasi-elastic scattering
of polarized electrons off the deuteron~\cite{Hasty00}, and
therefore influence the extraction from these data
(and those on the proton~\cite{Spayde00}) of the nucleon's
strange magnetic and axial form factors at a four-momentum
transfer squared of $0.1$ (GeV/c)$^2$.

The present is the first in a series of papers
dealing with the theoretical investigation of PV interaction
effects in two-nucleon systems: it is devoted to $p$$p$ elastic
scattering, and presents a calculation of the longitudinal
asymmetry induced by PV interactions in the lab-energy range
0--350 MeV.

The available experimental data on the $p$$p$ longitudinal
asymmetry is rather limited.  There are two measurements
at 15 MeV~\cite{Nagle79} and 45 MeV~\cite{Balzer84},
which yielded asymmetry values of $(-1.7\pm 0.8)\times 10^{-7}$ and
$(-2.3\pm 0.9)\times 10^{-7}$, respectively, as well as more
precise measurements at 13.6 MeV~\cite{Eversheim91},
45 MeV~\cite{Kistryn87}, and 221 MeV~\cite{Berdoz01} yielding
$(-0.95\pm 0.15)\times 10^{-7}$, 
$(-1.50\pm 0.23)\times 10^{-7}$, and
$(+0.84\pm 0.29)\times 10^{-7}$, 
respectively, and finally a measurement
at 800 MeV in Ref.~\cite{Yuan86}, which produced an asymmetry
value of $(+2.4\pm 1.1)\times 10^{-7}$.  
 
The theoretical (and, in fact, experimental) investigation
of PV effects induced by the weak interaction in the $p$$p$ system
began with the prediction by Simonius~\cite{Simonius72} that
the longitudinal asymmetry would have a broad maximum at energies
close to 50 MeV, and that, being dominated
by the $J$=0 partial waves, it would be essentially independent
of the scattering angle.  A number of theoretical studies of varying 
sophistication followed~\cite{Brown74,Henley75,Oka81}, culminating in the study 
by Driscoll and Miller~\cite{Driscoll89}, who used a distorted-wave
Born-approximation (DWBA) formulation of the PV scattering amplitude in
terms of exact wave functions obtained from solutions of
the Schr\"odinger equation with Coulomb and strong interactions. 
In fact, the authors of Ref.~\cite{Driscoll89} investigated the
sensitivity of the calculated asymmetry to a number of
realistic strong-interaction potentials constructed by the late 1980's.
The model adopted for the PV weak-interaction potential, however,
was that developed by Desplanques and collaborators~\cite{Desplanques80},
the so-called DDH model.  In the $p$$p$ sector, this potential
is parametrized in terms of $\rho$- and $\omega$-meson exchanges, in which
the PV $N$$N$$\rho$ and $N$$N$$\omega$ weak coupling constants 
are calculated in a quark model approach incorporating symmetry
techniques like SU(6)$_W$ and current algebra requirements.
Factoring in the limitations inherent to such an approach, however,
the authors of Ref.~\cite{Desplanques80} gave rather wide ranges
of uncertainty for these weak coupling constants.

The present work sharpens and updates that of Ref.~\cite{Driscoll89}. 
It adopts the DDH model for the PV weak-interaction potential, but
uses the latest generation of realistic, parity-conserving (PC)
strong-interaction potentials, the Argonne $v_{18}$~\cite{Wiringa95},
Nijmegen I~\cite{Stoks94}, and CD-Bonn~\cite{Machleidt01}.
Rather than employing the DWBA scheme of Ref.~\cite{Driscoll89}
to calculate the PV component of the $p$$p$ elastic scattering
amplitude, it solves the complete scattering problem in the presence
of these PC and PV potentials (including the Coulomb potential),
in either configuration- or momentum-space, depending on whether
the Argonne $v_{18}$ and Nijmegen I or CD-Bonn models are used.
Such an approach allows us to obtain the PC and PV wave functions
explicitly.  While this is unnecessary for the calculation reported 
here--the DWBA estimate, along the lines of Ref.~\cite{Driscoll89},
of the PV component of the $p$$p$ amplitude should suffice--it becomes
essential for the studies of $n$$p$ radiative
capture and deuteron electro-disintegration planned at a later stage.

The remainder of the present paper is organized as follows.
In Sec.~\ref{sec:pots} the PC and PV potentials used in this work
are briefly described, while in Sec.~\ref{sec:app} a self-consistent
treatment of the $p$$p$ scattering problem is provided along with a
discussion, patterned after that of Ref.~\cite{Driscoll89},
of the Coulomb contributions to the longitudinal asymmetries
measured in scattering and transmission experiments.
In Sec.~\ref{sec:res} the results for the asymmetry are presented;
in particular, their sentitivity to changes in the values of the
weak coupling constants and/or short-range cutoffs at the
strong- and weak-interaction vertices is studied.  Finally,
Sec.~\ref{sec:cons} contains some concluding remarks. 
\section{Parity-Conserving and Parity-Violating Potentials}
\label{sec:pots}

The parity-conserving (PC), strong-interaction potentials
used in the present work are the Argonne $v_{18}$ (AV18)~\cite{Wiringa95},
Nijmegen I (NIJ-I)~\cite{Stoks94}, and CD-Bonn (BONN)~\cite{Machleidt01}
models.  The AV18 and NIJ-I potentials were fitted to the Nijmegen
database of 1992~\cite{Bergervoet90,Stoks93}, consisting of 1787 $p$$p$
data, and both produced $\chi^2$ per datum close to one.  The latest version
of the charge-dependent Bonn potential, however, has been fit to the
1999 database, consisting of 2932 $p$$p$ data, for which it gives a $\chi^2$ 
per datum of 1.01~\cite{Machleidt01}.  The substantial increase in
the number of $p$$p$ data is due to the development of novel
experimental techniques--internally polarized gas targets and stored, cooled
beams.  Indeed, using this technology, IUCF has produced a large
number of $p$$p$ spin-correlation parameters of very high precision,
see for example Ref.~\cite{Przewoski98}.  It is worth noting that
the AV18 potential, as an example, fits the post-1992 and both pre- and
post-1992 $p$$p$ data with $\chi^2$'s of 1.74 and 1.35,
respectively~\cite{Machleidt01}.  Therefore, while the quality of
their fits has deteriorated somewhat in regard to the 
extended 1999-database, the AV18 and NIJ-I models can still be
considered \lq\lq realistic\rq\rq.

These realistic potentials consist of a long-range part due to
one-pion exchange (OPE), and a short-range part either modeled by
one-boson exchange (OBE), as in the BONN and NIJ-I models, or
parameterized in terms of suitable functions of two-pion range
or shorter, as in the AV18 model.  While these potentials are (almost)
phase-equivalent, they differ in the treatment of non-localities.
AV18 is local (in $LSJ$ channels), while BONN and NIJ-I have
strong non-localities.  In particular, BONN has a non-local
OPE component.  However, it has been known
for some time~\cite{Friar77}, and recently
re-emphasized in Ref.~\cite{Forest00}, that the
local and non-local OPE terms are related to each other
by a unitary transformation.  Therefore, the differences between
local and non-local OPE cannot be of any consequence for the prediction
of observables, such as binding energies or electromagnetic form
factors, provided, of course, that three-body interactions and/or
two-body currents generated by the unitary transformation are
also included~\cite{Coon86}.  This fact has been
demonstrated~\cite{Schiavilla01} in a calculation of
the deuteron structure function $A(q)$ and tensor observable $T_{20}(q)$,
based on the local AV18 and non-local BONN 
models and associated (unitarily consistent) electromagnetic currents.
The remaining small differences between the calculated $A(q)$ and
$T_{20}(q)$ are due to the additional short-range non-localities
present in the BONN model.  Therefore, provided
that consistent calculations--in the sense above--are performed,
present \lq\lq realistic\rq\rq potentials will lead to very similar
predictions for nuclear observables, at least to the extent that
these are influenced predominantly by the OPE component.

As already mentioned in Sec.~\ref{sec:intro}, the form of
the parity-violating (PV) weak-interaction potential was derived
in Ref.~\cite{Desplanques80}--the DDH model:

\begin{eqnarray}
v^{\rm PV}=\sum_{\alpha=\rho,\omega}&-&
\frac{g_\alpha \, h^{pp}_\alpha}{4\pi} \frac{m_\alpha}{m}
\Big\{ m_\alpha (1+\kappa_\alpha)Y^\prime (m_\alpha r)
({\bbox \sigma}_1\times {\bbox \sigma}_2) \cdot \hat{\bf r} \nonumber \\
&+&({\bbox \sigma}_1 - {\bbox \sigma}_2) \cdot \left[ {\bf p} \, , 
\, Y(m_\alpha r)\right]_{+} \Big\} \>\>,
\label{eq:DDH}
\end{eqnarray}
where the relative position and momentum are defined
as ${\bf r}={\bf r}_1-{\bf r}_2$ and ${\bf p}=({\bf p}_1-{\bf p}_2)/2$,
respectively, $\left[ \dots \, , \, \dots \right]_{+}$ denotes
the anticommutator, and $m$ and $m_\alpha$ are the proton and vector-meson
($\rho$ or $\omega$) masses, respectively.  Note that
the first term in Eq.~(\ref{eq:DDH}) is usually written
in the form of a commutator, since 

\begin{equation}
{\rm i} \left[ {\bf p} \, , \, Y(m_\alpha r) \right]_{-}
=m_\alpha Y^\prime (m_\alpha r)\, \hat{\bf r} \>\>\>,
\end{equation}
where $Y^\prime(x)$ denotes its derivative ${\rm d}Y(x)/{\rm d}x$.
The Yukawa function
$Y(x_\alpha)$, suitably modified by the inclusion of
monopole form factors, is given by

\begin{equation}
Y(x_\alpha)= \frac{1}{x_\alpha}\Bigg\{ {\rm e}^{-x_\alpha}
-{\rm e}^{(\Lambda_\alpha/m_\alpha) x_\alpha} \left[ 1+
\frac{1}{2}\frac{\Lambda_\alpha}{m_\alpha}
\left( 1-\frac{m^2_\alpha}{\Lambda^2_\alpha} \right)
x_\alpha \right] \Bigg\} \>\>,
\end{equation}
where $x_\alpha \equiv m_\alpha r$.  Finally, the values
for the strong-interaction $\rho$- and $\omega$-meson vector
and tensor coupling constants $g_\alpha$ and $\kappa_\alpha$, as well as
for the cutoff parameters $\Lambda_\alpha$, are
taken from the BONN model~\cite{Machleidt01}, and
are listed in Table~\ref{tb:gs}.  The weak-interaction
coupling constants $h^{pp}_\rho$ and $h^{pp}_\omega$
correspond to the following combinations of DDH parameters

\begin{eqnarray}
h^{pp}_\rho&=&h_{\rho_0}+h_{\rho_1}+ \frac{h_{\rho_2}}{\sqrt{6}} \>\>, \\
h^{pp}_\omega&=&h_{\omega_0}+h_{\omega_1} \>\>.
\end{eqnarray}
Their values, reported in Table~\ref{tb:gs} in the column labeled
(DDH-adj), are obtained by
fitting the available data on the longitudinal asymmetry, see Sec.~\ref{sec:res}.
The values corresponding to the \lq\lq best\rq\rq estimates
for the $h_{\rho_i}$ and $h_{\omega_i}$ suggested in
Ref.~\cite{Desplanques80} are also listed in Table~\ref{tb:gs}
in the column (DDH-orig).
Indeed, one of the goals of the present work is to
study the sensitivity of the calculated longitudinal asymmetry
to variations in both the PV coupling constants and cutoff
parameters.  In this respect, it should also be noted that,
in the limit $\Lambda_\rho$=$\Lambda_\omega$ and ignoring the
small mass difference between $m_\rho$ and $m_\omega$, were it not
for the different values of the tensor couplings $\kappa_\rho$ and
$\kappa_\omega$, the $\rho$- and $\omega$-meson terms
in $v^{\rm PV}$ would collapse to a single term of strength
proportional to $g_\rho h^{pp}_\rho + g_\omega h^{pp}_\omega$.
\section{Formalism}
\label{sec:app}

In this section we discuss the $p$$p$ scattering problem 
in the presence of a potential $\overline{v}$ given by

\begin{equation}
\overline{v}= v^{\rm PC}+v^{\rm PV}+v^{\rm C} \>\>,
\end{equation}
where $v^{\rm PC}$ and $v^{\rm PV}$ denote the parity-conserving
and parity-violating components induced by the strong and weak
interactions, respectively, and $v^{\rm C}$ is the Coulomb
potential.
 
\subsection{Partial-Wave Expansions of Scattering State, $T$- and $S$-Matrices}
\label{sec:pwest}

The Lippmann-Schwinger equation for the $p$$p$ scattering
state $\mid{\bf p},SM_S\rangle^{(\pm)}$, where ${\bf p}$
is the relative momentum and $SM_S$ specifies
the spin state, can be written as~\cite{Goldberger64}

\begin{equation}
\mid{\bf p},SM_S\rangle^{(\pm)}=
\mid {\bf p},SM_S\rangle^{(\pm)}_{\rm C} +
\frac{1}{E-H_0-v^{\rm C}\pm {\rm i}\epsilon}\, v\,
\mid{\bf p},SM_S\rangle^{(\pm)} \>\>,
\label{eq:LS}
\end{equation}
where $H_0$ is the free Hamiltonian, $v=v^{\rm PC}+v^{\rm PV}$, and
$\mid \dots\rangle^{(\pm)}_{\rm C}$ are the eigenstates of
$H_0+v^{\rm C}$,

\begin{equation}
\left( E-H_0-v^{\rm C} \right)
\mid{\bf p},SM_S\rangle^{(\pm)}_{\rm C}=0 \>\>,
\end{equation}
with wave functions given by

\begin{eqnarray}
\phi^{(\pm)}_{ {\bf p}, SM_S}({\bf r})&=&
\langle {\bf r}\mid {\bf p},SM_S\rangle^{(\pm)}_{\rm C} \nonumber \\
&=&4\pi\sqrt{2} \sum_{JM_JL} {\rm i}^L\, \epsilon_{LS} \,\,
{\rm e}^{\pm {\rm i}\sigma_L} \frac{F_L(\eta;pr)}{pr} 
[Z_{LSM_S}^{JM_J}(\hat{\bf p})]^* \,
{\cal Y}_{LSJ}^{M_J}(\hat{\bf r}) \>\>.
\label{eq:pwe}
\end{eqnarray}
Here $F_L(\eta;\rho)$ denotes the regular Coulomb
wave function~\cite{Abramowitz74}, while the parameter $\eta$
and Coulomb phase-shift $\sigma_L$ are given by

\begin{eqnarray}
\eta &=& \alpha \mu/p \>\>, \\
\sigma_L&=& {\rm arg}\left[\Gamma(L+1+{\rm i} \eta)\right] \>\>,
\end{eqnarray}
where $\alpha$ is the fine structure constant and $\mu$ is the
reduced mass.  Finally, the following definitions have also
been introduced:

\begin{equation}
Z_{LSM_S}^{JM_J}(\hat{\bf p})\equiv
\sum_{M_L} \langle LM_L,SM_S\mid JM_J\rangle
\, Y_{LM_L}(\hat {\bf p}) \>\>,
\end{equation}
\begin{equation}
\epsilon_{LS} \equiv \frac{1}{2} \Big[1 +(-1)^{L+S} \Big] \>\>.
\end{equation}
The factor $\epsilon_{LS}$ ensures that the wave functions $\phi^{(\pm)}$
are properly antisymmetrized.  Note that in the limit $\eta$=0, equivalent
to ignoring the Coulomb potential, the latter reduce to (antisymmetrized)
plane waves, 

\begin{equation}
\phi^{(\pm)}_{ {\bf p}, SM_S}({\bf r}) \rightarrow \frac{1}{\sqrt{2}}
\Big[{\rm e}^{ {\rm i} {\bf p} \cdot {\bf r}} +(-)^{S} 
     {\rm e}^{-{\rm i} {\bf p} \cdot {\bf r}}\Big] 
\chi^S_{M_S} \>\>. 
\end{equation}

The $\overline{T}$-matrix corresponding
to the potential $v+v^{\rm C}$ can be expressed as~\cite{Goldberger64}

\begin{equation}
\overline{T}({\bf p}^\prime,S^\prime M_S^\prime;{\bf p},SM_S)=
T({\bf p}^\prime,S^\prime M_S^\prime;{\bf p},SM_S)+
T^{\rm C}({\bf p}^\prime,S^\prime M_S^\prime;{\bf p},SM_S) \>\>,
\label{eq:tmas}
\end{equation}
where $T^{\rm C}$ is the (known) $T$-matrix corresponding only
to the Coulomb potential~\cite{Goldberger64}, and

\begin{equation}
T({\bf p}^\prime,S^\prime M_S^\prime;{\bf p},SM_S)=
^{(-)}_{\rm C}\!\!\langle {\bf p}^\prime,S^\prime M_S^\prime
\mid T \mid {\bf p},SM_S\rangle^{(+)} \>\>.
\end{equation}
Insertion of the complete set of states
$\mid {\bf p},SM_S\rangle^{(-)}_{\rm C}$ into the
right-hand-side of the Lippmann-Schwinger equation leads to

\begin{equation}
\mid{\bf p},SM_S\rangle^{(+)}=\mid{\bf p},SM_S\rangle^{(+)}_{\rm C}
+\sum_{S^\prime M_S^\prime}\int\frac{{\rm d}{\bf p}^\prime}{(2\pi)^3}
\frac{1}{2} \mid{\bf p}^\prime,S^\prime M_S^\prime\rangle^{(-)}_{\rm C}
\frac{T({\bf p}^\prime,S^\prime M_S^\prime;{\bf p},SM_S)}
{E-p^{\prime 2}/(2\mu) +{\rm i}\epsilon} \>\>,
\label{eq:LSa}
\end{equation}
from which the partial wave expansion of the scattering state
is easily obtained by first noting that
the potential, and hence the $T$-matrix, can be expanded as

\begin{eqnarray}
^{(-)}_{\rm C}\!\langle {\bf p}^\prime,S^\prime M_S^\prime
\mid v \mid {\bf p},SM_S\rangle^{(+)}_{\rm C} &=& 2(4\pi)^2\! \sum_{JM_J}
\sum_{L L^\prime}
\epsilon_{L^\prime S^\prime}\,\, \epsilon_{LS} \,
{\rm e}^{{\rm i}\sigma_{L^\prime}}
{\rm e}^{{\rm i}\sigma_L}
Z_{L^\prime S^\prime M_S^\prime}^{JM_J}(\hat{\bf p}^\prime) \nonumber \\
&&[Z_{LSM_S}^{JM_J}(\hat{\bf p})]^*\,
v^J_{L^\prime S^\prime,LS} (p^\prime;p) \>\>,
\label{eq:tmae}
\end{eqnarray}
with
\begin{equation}
v^J_{L^\prime S^\prime,LS}(p^\prime;p)
={\rm i}^{L-L^\prime}\int {\rm d}{\bf r}
\frac{F_{L^\prime}(\eta;p^\prime r)}{p^\prime r}
{\cal Y}_{L^\prime S^\prime J}^{M_J \dagger}
v({\bf r}) 
{\cal Y}_{LSJ}^{M_J} 
\frac{F_L(\eta;p r)}{p r} \>\>.
\end{equation}
After insertion of the corresponding expansion for the $T$-matrix
into Eq.~(\ref{eq:LSa}) and a number of standard manipulations,
the scattering-state wave function can be written as

\begin{equation}
\psi^{(+)}_{ {\bf p}, SM_S}({\bf r})=
4\pi\sqrt{2} \sum_{JM_J}\, \sum_{L L^\prime S^\prime}
{\rm i}^{L^\prime}\, \epsilon_{L^\prime S\prime}
\, \epsilon_{LS} \, {\rm e}^{{\rm i}\sigma_L}
\, [Z_{LSM_S}^{JM_J}(\hat{\bf p})]^*
\frac{w^J_{L^\prime S^\prime,LS}(r;p)}{r}
{\cal Y}_{L^\prime S^\prime J}^{M_J}(\hat{\bf r}) \>\>,
\end{equation}
with

\begin{eqnarray}
\frac{w^J_{L^\prime S^\prime,LS}(r;p)}{r}&=&
\Bigg[ \delta_{L,L^\prime}\delta_{S,S^\prime}
\frac{F_{L^\prime}(\eta;pr)}{pr} \nonumber \\
&+&\frac{2}{\pi}\int_0^\infty {\rm d}p^\prime p^{\prime 2}
\frac{F_{L^\prime}(\eta;p^\prime r)}{p^\prime r}\frac{1}
{E-p^{\prime 2}/(2\mu)+{\rm i}\epsilon}
T^J_{L^\prime S^\prime,LS}(p^\prime;p) \Bigg] \>\>.
\end{eqnarray}
The (complex) radial wave function $w(r)$ behaves in the
asymptotic region $r \rightarrow \infty$ as

\begin{equation}
\frac{w^J_{\alpha^\prime,\alpha}(r;p)}{r}\simeq \frac{1}{2} \Big[
\delta_{\alpha,\alpha^\prime} h^{(2)}_{L^\prime}(\eta;pr)
+h^{(1)}_{L^\prime}(\eta;pr) S^J_{\alpha^\prime , \alpha}(p) \Big] \>\>,
\label{eq:asy}
\end{equation}
where the label $\alpha$ ($\alpha^\prime$) stands for the set of quantum numbers
$LS$ ($L^\prime S^\prime$), the on-shell ($p^\prime=p$) $S$-matrix
has been introduced,

\begin{equation}
S^J_{\alpha^\prime , \alpha}(p)=\delta_{\alpha,\alpha^\prime}
-4{\rm i}\, \mu p \, T^J_{\alpha^\prime , \alpha} (p;p) \>\>,
\label{eq:sma}
\end{equation}
and the functions $h^{(1,2)}(\eta;\rho)$ are defined in terms
of the regular and irregular ($G_L$) Coulomb functions as

\begin{equation}
h^{(1,2)}_L(\eta;\rho)=\frac{F_L(\eta;\rho)\mp {\rm i}\, 
G_L(\eta;\rho)}{\rho} \>\>.
\end{equation}
Again, in the limit $\eta =0$, $F_L(\eta;\rho)/\rho \rightarrow
j_L(\rho)$ and $G_L(\eta;\rho)/\rho \rightarrow -n_L(\rho)$, where
$j_L(\rho)$ and $n_L(\rho)$ are the spherical Bessel functions,
and the familiar expressions for the partial wave-expansion
of the scattering state, $S$- and $T$-matrices are
recovered~\cite{Goldberger64}. 

\subsection{Schr\"odinger Equation, Phase-Shifts, and Mixing Angles}
\label{sec:phase}

The coupled-channel Schr\"odinger equations for the radial wave
functions $w(r)$ read:

\begin{equation}
\Bigg[ -\frac{ {\rm d}^2}{{\rm d}r^2} + \frac{L(L+1)}{r^2}
-p^2 \Bigg] w^J_{\alpha^\prime,\alpha}(r;p)
+\sum_{\beta} r \, v^J_{\alpha^\prime,\beta}(r)\, \frac{1}{r}
w^J_{\beta,\alpha}(r;p) = 0  \>\>,
\label{eq:cschr}
\end{equation}
with

\begin{equation}
v^J_{\alpha^\prime,\alpha}(r)={\rm i}^{L-L^\prime}\, 2\mu \,
\int{\rm d}\Omega\, {\cal Y}_{\alpha^\prime J}^{M_J \dagger}
\, v({\bf r})\, {\cal Y}_{\alpha J}^{M_J} \>\>,
\label{eq:vpe}
\end{equation}
where, because of time reversal invariance, the matrix
$v^J_{\alpha^\prime,\alpha}$ can be shown to be
real and symmetric (this is the reason for
the somewhat unconventional phase factor
in Eq.~(\ref{eq:vpe}); in order to mantain symmetry for
both the $v^{\rm PC}$- and $v^{\rm PV}$-matrices, and
hence the $S$-matrix, the states used here differ by a factor ${\rm i}^L$ from
those usually used in nucleon-nucleon scattering analyses).
The asymptotic behavior of the $w(r)$'s is given 
in Eq.~(\ref{eq:asy}).

The Pauli principle requires that there be a single channel
when $J$ is odd, and three coupled channels when $J$ is even,
with the exception of $J$=0 in which case there are only two 
coupled channels, $^1$S$_0$ and $^3$P$_0$.  The situation
is summarized in Table~\ref{tb:chan}.
Again because of the invariance under time-inversion
transformations of $v^{\rm PC}+v^{\rm PV}$,
the $S$-matrix is symmetric (apart from also
being unitary), and can therefore be written, for the coupled
channels having $J$ even, as~\cite{Goldberger64}

\begin{equation}
S^J = U^{\rm T} \, S^J_{\rm D} \, U \>\>,
\label{eq:sdiag}
\end{equation}
where $U$ is a real orthogonal matrix, and $S^J_D$ is a diagonal
matrix of the form

\begin{equation}
S^J_{\rm D;\alpha^\prime,\alpha} = \delta_{\alpha^\prime,\alpha}
{\rm e}^{2 {\rm i} \delta^J_\alpha} \>\>.
\end{equation}
Here $\delta^J_\alpha$ is the (real) phase-shift in
channel $\alpha$, which is function of the energy $p=\sqrt{2\mu\, E}$.
The mixing matrix $U$ can be written as

\begin{eqnarray}
U &=& U^{(12)} \quad \qquad \qquad  J=0 \>\>, \\
  &=& \prod_{1 \leq i < j \leq 3} U^{(ij)}
\qquad J \geq 2\>\>{\rm with}\>\> J \>\>{\rm even}\>\>,
\label{eq:uma}
\end{eqnarray}
where $U^{(ij)}$ is the $2 \times 2$ or $3 \times 3$
orthogonal matrix, that includes the coupling between
channels $i$ and $j$ only, for example

\noindent
\centerline{
$
  U^{(13)}= \left[ \begin{array}{ccc}
           {\rm cos}\,\epsilon^J_{13} & 0 & {\rm sin}\,\epsilon^J_{13} \\
           0                        & 1 & 0                        \\
          -{\rm sin}\,\epsilon^J_{13} & 0 & {\rm cos}\,\epsilon^J_{13} 
           \end{array} \right]
\simeq 1 +\epsilon^J_{13} \left[ \begin{array}{ccc}
           0 & 0 & 1 \\
           0 & 0 & 0 \\
          -1 & 0 & 0 \\
\end{array} \right ] \>\>.
$ }
\vskip 0.7cm

\noindent
Thus, for $J$=0 there are two phase-shifts and a mixing
angle, while for $J$ even $\geq 2$ there are
three phases and three mixing angles.  Of course, since
$| v^{\rm PV} | \ll | v^{\rm PC} |$, the mixing angles $\epsilon^J_{ij}$
induced by $v^{\rm PV}$ are $\ll 1$, a fact already exploited in the
last expression above for $U$.  Given the channel ordering in
Table~\ref{tb:chan}, Table~\ref{tb:mixing} specifies
which of the channel mixings are induced by $v^{\rm PC}$
and which by $v^{\rm PV}$. 

The reality of the potential matrix elements
$v^J_{\alpha^\prime,\alpha}(r)$ makes
it possible to construct real solutions of the
Schr\"odinger equation~(\ref{eq:cschr}).
The problem is reduced to determining the
relation between these solutions and the complex $w(r)$'s functions.
Using Eq.~(\ref{eq:sdiag}) and $U^{\rm T} U=1$,
the $w(r)$'s can be expressed in the asymptotic region as

\begin{eqnarray}
\frac{w^J_{\alpha^\prime,\alpha}}{r} &\simeq&
\sum_\beta (U^{\rm T})_{\alpha^\prime \beta} \, {\rm e}^{ {\rm i}\delta^J_\beta}
\frac{ h^{(2)}_{\alpha^\prime} {\rm e}^{-{\rm i}\delta^J_\beta}
+  h^{(1)}_{\alpha^\prime} {\rm e}^{ {\rm i}\delta^J_\beta} }{2}
U_{\beta \alpha} \nonumber \\
&=& \sum_\beta (U^{\rm T})_{\alpha^\prime \beta} \,
{\rm e}^{ {\rm i}\delta^J_\beta} \,
\frac{ {\rm sin}[pr-L^\prime \, \pi/2 
-\eta {\rm lg}(2pr)+\sigma_{L^\prime}+\delta^J_\beta] }{pr}
\, U_{\beta \alpha} \>\>,
\end{eqnarray}
where the $\epsilon_{L^\prime S^\prime}$ has been
dropped for simplicity.  The expression above is real apart from
the ${\rm exp}({\rm i}\delta^J_\beta)$.  To eliminate
this factor, the following linear combinations of the
$w(r)$'s are introduced

\begin{eqnarray}
\frac{u^J_{\alpha^\prime,\alpha}}{r} &\equiv& \sum_\beta
{\rm e}^{-{\rm i}\delta^J_\beta}
\frac{w^J_{\alpha^\prime,\beta}}{r}(U^{\rm T})_{\beta \alpha} \nonumber \\
&\simeq&(U^{\rm T})_{\alpha^\prime \alpha}
\frac{{\rm cos}\, \delta^J_\alpha \, F_{L^\prime}(\eta;pr)
     +{\rm sin}\, \delta^J_\alpha \, G_{L^\prime}(\eta;pr)}{pr} \>\>,
\label{eq:uasy}
\end{eqnarray}
and the $u(r)$'s are then the sought real solutions of Eq.~(\ref{eq:cschr}).

The asymptotic behavior of the $u(r)$'s can now be read off
from Eq.~(\ref{eq:uasy}) once the $U$-matrices above have been
constructed.  The latter can be written as, up
to linear terms in the \lq\lq small\rq\rq mixing angles induced
by $v^{\rm PV}$,

\noindent
\centerline{
$
  U= \left[ \begin{array}{cc}
            1             & \epsilon^0_{12} \\
           -\epsilon^0_{12} & 1
           \end{array} \right] \quad J=0 \>\>,
$}
\vskip 0.7cm
\centerline{
$
  U= \left[ \begin{array}{ccc}
    {\rm cos}\epsilon^J_{12}
  & {\rm sin}\epsilon^J_{12}
  &  \epsilon^J_{13}{\rm cos}\epsilon^J_{12}+\epsilon^J_{23}{\rm sin}\epsilon^J_{12} \\
   -{\rm sin}\epsilon^J_{12}
  & {\rm cos}\epsilon^J_{12}
  & -\epsilon^J_{13}{\rm sin}\epsilon^J_{12}+\epsilon^J_{23}{\rm cos}\epsilon^J_{12} \\
  -\epsilon^J_{13} & -\epsilon^J_{23} & 1 \\
\end{array} \right ]  \quad J \geq 2 ,\>\> J\>\>{\rm even} \>\>.
$}
\vskip 0.7cm

Inverting the first line of Eq.~(\ref{eq:uasy}),

\begin{equation}
\frac{w^J_{\alpha^\prime,\alpha}}{r} = \sum_\beta
{\rm e}^{ {\rm i}\delta^J_\beta}
\frac{u^J_{\alpha^\prime,\beta}}{r}\,U_{\beta \alpha} \>\>,
\end{equation}
and inserting the resulting expressions into Eq.~(\ref{eq:cschr})
leads to the (in general, coupled-channel) Schr\"odinger equations
satisfied by the (real) functions $u(r)$.  They are identical to
those of Eq.~(\ref{eq:cschr}), but for the $w(r)$'s being replaced
by the $u(r)$'s.  These equations are then solved by standard numerical
techniques.  Note that: i) $v^J_{\alpha^\prime,\alpha} =
v^{J,\, {\rm PC}}_{\alpha^\prime,\alpha}$, since the diagonal
matrix elements of $v^{\rm PV}$ vanish because of parity selection rules;
ii) in the coupled equations with $J$ even, terms of the type
$r \,v^{J,\, {\rm PV}}_{\alpha^\prime,\beta}(r) u^J_{\beta,\alpha}(r)/r$
involving the product of a parity-violating potential matrix element
with a $v^{\rm PV}$-induced wave function are neglected.

\subsection{Amplitudes, Cross Sections, and the Parity-Violating Asymmetry}

The amplitude for $p$$p$ elastic scattering from an initial
state with spin projections $m_1$, $m_2$ to a final state
with spin projections $m^\prime_1$, $m^\prime_2$ is given
by

\begin{eqnarray}
\langle m_1^\prime m_2^\prime \mid \overline{M} \mid m_1 m_2\rangle&=&
\sum_{S^\prime M_S^\prime,SM_S}
\langle \frac{1}{2} m_1^\prime ,
        \frac{1}{2} m_2^\prime \mid S^\prime M_S^\prime\rangle
\langle \frac{1}{2} m_1 ,\frac{1}{2} m_2 \mid S M_S\rangle \nonumber \\
&&\overline{M}_{S^\prime M_S^\prime, SM_S}(E,\theta) \>\>,
\end{eqnarray}
where the amplitude $\overline{M}$ is related to the
$\overline{T}$-matrix defined in Eq.~(\ref{eq:tmas}) via

\begin{equation}
\overline{M}_{S^\prime M_S^\prime, SM_S}(E,\theta) =-\frac{\mu}{2\pi}
\overline{T}({\bf p}^\prime,S^\prime M_S^\prime;p\hat{\bf z},SM_S) \>\>.
\end{equation}
Note that the direction of the initial momentum
${\bf p}$ has been taken to define the spin quantization axis
(the $z$-axis), $\theta$ is the angle between
$\hat {\bf p}$ and $\hat{\bf p}^\prime$, the direction
of the final momentum, and the energy $E=p^2/(2\mu)$
($= p^{\prime 2}/(2\mu)$).  The amplitude $\overline{M}$
is split into two terms, $\overline{M}=M+M^{\rm C}$,
as in Eq.~(\ref{eq:tmas}).  Using the expansion of
the $T$-matrix, Eq.~(\ref{eq:tmae}) with $v^J_{L^\prime S^\prime,LS}$
replaced by $T^J_{L^\prime S^\prime,LS}$, and the relation
between the $S$- and $T$-matrices, Eq.~(\ref{eq:sma}), the amplitude induced
by $v^{\rm PC}+v^{\rm PV}$ can be expressed as

\begin{eqnarray}
M_{S^\prime M_S^\prime, SM_S}(E,\theta)&=& \sqrt{4\pi} \sum_{JLL^\prime}
\sqrt{2L+1}\, \epsilon_{L^\prime S^\prime}\, \epsilon_{LS} \,
\langle L^\prime (M_S-M^\prime_S), S^\prime M^\prime_S\mid J M_S\rangle \nonumber \\
&&\langle L 0, SM_S\mid J M_S\rangle Y_{L^\prime (M_S-M_S^\prime)}(\theta)\,
{\rm e}^{{\rm i}\sigma_{L^\prime}}\,
\frac{S^J_{L^\prime S^\prime,LS}(p) - \delta_{L^\prime,L}
\,\delta_{S^\prime,S}}{{\rm i} p}
\,{\rm e}^{{\rm i}\sigma_L} \>\>,
\label{eq:am}
\end{eqnarray}
while the partial-wave expansion of the amplitude
associated with the Coulomb potential reads~\cite{Goldberger64}: 

\begin{equation}
M^{\rm C}_{S^\prime M_S^\prime, SM_S}(E,\theta)=\delta_{S^\prime,S} \,
\delta_{M^\prime_S,M_S} \,\sqrt{4\pi} \sum_{L} \sqrt{2L+1}\,\epsilon_{LS}\,
Y_{L0}(\theta) \frac{ {\rm e}^{2 {\rm i}\sigma_L}-1}{{\rm i} p}\>\>.
\label{eq:amc}
\end{equation}

The differential cross section for scattering of a proton with initial
polarization $m_1$ is then given by

\begin{equation}
\overline{\sigma}_{m_1}(E,\theta)=\frac{1}{2} \sum_{m_2} \sum_{m_1^\prime m_2^\prime}
\mid \langle m_1^\prime m_2^\prime \mid \overline{M} \mid m_1 m_2\rangle\mid^2 \>\>,
\end{equation}
and the longitudinal asymmetry is defined as

\begin{equation}
\overline{A}(E,\theta)=\frac{\overline{\sigma}_{+}(E,\theta)
 - \overline{\sigma}_{-}(E,\theta)}
 {\overline{\sigma}_{+}(E,\theta) + \overline{\sigma}_{-}(E,\theta) } \>\>,
\end{equation}
where $\pm$ denote the initial polarizations $\pm 1/2$.  Carrying out
the spin sums leads to the following expression for the asymmetry: 

\begin{equation}
\overline{A}(E,\theta)=\frac{
\sum_{S^\prime M^\prime_S} \left[
\overline{M}_{S^\prime M_S^\prime, 00}(E,\theta)
\overline{M}^*_{S^\prime M_S^\prime, 10}(E,\theta) + {\rm c.c.}\right] }
{\sum_{S^\prime M^\prime_S} \sum_{S M_S} 
\mid \overline{M}_{S^\prime M_S^\prime, S M_S}(E,\theta) \mid^2 }\>\>,
\label{eq:angc}
\end{equation}
from which it is clear that the numerator would vanish
in the absence of parity-violating interactions, since
$v^{\rm PC}+v^{\rm C}$, in contrast to $v^{\rm PV}$,
cannot change the total spin $S$ of the $p$$p$ pair.

Parity-violating scattering experiments typically measure
the asymmetry weighted over a range $[\theta_1,\theta_2]$
of scattering angles, 

\begin{equation}
\langle\, \overline{A}(E)\, \rangle =
\frac{\int_{\theta_1 \leq \theta \leq \theta_2}
{\rm d}\Omega\, \overline{\sigma}(E,\theta) \, \overline{A}(E,\theta) }
{\int_{\theta_1 \leq \theta \leq \theta_2}{\rm d}\Omega \,
 \overline{\sigma}(E,\theta)} \>\>,
\end{equation}
where $\overline{\sigma}=
\left( \overline{\sigma}_{+}+\overline{\sigma}_{-} \right)/2$
is the spin-averaged differential cross section.
In contrast, transmission experiments measure the
transmission of a polarized proton beam through a target.  A cross
section is then inferred from the transmission measurement.  Beam
particles elastically scattered by angles greater than some
small critical angle $\theta_0$ are removed from the beam, thus
reducing the observed transmission and adding to the inferred
cross section.  Beam particles scattered at angles smaller than
$\theta_0$ are not distinguished from the beam and do not
contribute to the cross section.  To derive an expression
for the asymmetry in this case, one needs to carefully
consider the Coulomb contribution to the cross section--a divergent
quantity in the limit $\theta_0 \rightarrow 0$.  To this end,
following Ref.~\cite{Holdeman65}, one first defines the 
differential cross sections

\begin{eqnarray}
\sigma_{S^\prime M^\prime_S,SM_S}(E,\theta) &\equiv& 
 \mid \overline{M}_{S^\prime M^\prime_S,SM_S}(E,\theta) \mid^2
-\mid M^{\rm C}_{S^\prime M^\prime_S,SM_S}(E,\theta) \mid^2 \nonumber \\
&=&\mid M_{S^\prime M^\prime_S,SM_S}(E,\theta) \mid^2+
2\,\Re\Big[M_{S^\prime M^\prime_S,SM_S}(E,\theta)
          M^{{\rm C}\, *}_{S^\prime M^\prime_S,SM_S}(E,\theta)\Big]\>\>,
\end{eqnarray}
and 

\begin{equation}
\sigma^{\rm C}_{S^\prime M^\prime_S,SM_S}(E,\theta)=
\mid M^{\rm C}_{S^\prime M^\prime_S,SM_S}(E,\theta) \mid^2 \>\>,
\end{equation}
and hence

\begin{equation}
\overline{\sigma}_{S^\prime M^\prime_S,SM_S}(E,\theta)=
\sigma_{S^\prime M^\prime_S,SM_S}(E,\theta)+
\sigma^{\rm C}_{S^\prime M^\prime_S,SM_S}(E,\theta) \>\>.
\end{equation}
In transmission experiments, the quantity of interest is

\begin{eqnarray}
\overline{\sigma}_{SM_S,>}(E)&\equiv&
2\pi \int_{\theta_0}^{\pi/2}{\rm d}\theta \,{\rm sin}\theta 
\sum_{S^\prime M^\prime_S}
\sigma_{S^\prime M^\prime_S,SM_S}(E,\theta) \nonumber \\
&=&\sigma_{SM_S,>}(E)+\sigma^{\rm C}_{SM_S,>}(E) \>\>,
\label{eq:s_tra}
\end{eqnarray}
where $\sigma^{\rm C}_{SM_S,>}(E)$ is explicitly given by

\begin{equation}
\sigma^{\rm C}_{SM_S,>}(E)=\pi \left(\frac{\eta}{p}\right)^2
\Bigg[ \frac{1}{{\rm sin}^2\theta_0/2} - \frac{1}{{\rm cos}^2\theta_0/2}
-\frac{(-)^S}{\eta} {\rm sin}
\left[ 2\eta\, {\rm lg}\, {\rm tg}(\theta_0/2)\right]\Bigg] \>\>.
\label{eq:scc}
\end{equation}
To evaluate $\sigma_{SM_S,>}(E)$, one writes, following Ref.~\cite{Holdeman65}:

\begin{eqnarray}
\sigma_{SM_S,>}(E)&=&\sigma_{SM_S}(E)-
2\pi\int_{\epsilon \rightarrow 0}^{\theta_0}{\rm d}\theta \,
{\rm sin}\theta \, \sum_{S^\prime M^\prime_S}
 \Bigg[ \mid M_{S^\prime M^\prime_S,SM_S}(E,\theta) \mid^2 \nonumber \\
&+&2\,\Re\Big[M_{S^\prime M^\prime_S,SM_S}(E,\theta)
          M^{{\rm C}\, *}_{S^\prime M^\prime_S,SM_S}(E,\theta)\Big] \Bigg] \>\>.
\label{eq:sbc}
\end{eqnarray}
Application of the optical theorem to the total cross sections
$\overline{\sigma}$ and $\sigma^{\rm C}$ allows
one to deduce

\begin{equation}
\sigma_{SM_S}(E)=\overline{\sigma}_{SM_S}(E)-\sigma^{\rm C}_{SM_S}(E)
=\frac{4\pi}{p}\Im \Big[ M_{SM_S,SM_S}(E,0) \Big] \>\>,
\end{equation}
and the determination of the cross section $\sigma_{SM_S,>}(E)$ is
reduced to evaluating the integral on the right-hand-side of Eq.~(\ref{eq:sbc}).
For sufficiently small $\theta_0$ and by appropriately taking the limit
$\epsilon \rightarrow 0$ in the integral of the interference
term $MM^{{\rm C}\, *}$, which essentially entails taking the limit
term by term in the partial-wave expansion of $M^{\rm C}$, one finds

\begin{equation}
\sigma_{SM_S,>}(E)=\frac{4\pi}{p} \Im \Bigg[ M_{SM_S,SM_S}(E,0)
{\rm e}^{2{\rm i}[ \eta \, {\rm lg}\, {\rm sin}(\theta_0/2)-\sigma_0]}\Bigg] \>\>,
\label{eq:sc}
\end{equation}
neglecting terms of order $\theta_0^2$ and higher.  Using
Eqs.~(\ref{eq:scc}) and~(\ref{eq:sc}), the longitudinal asymmetry
measured in transmission experiments is obtained as

\begin{equation} 
\overline{A}_{>}(E)=
\frac{\Im \Big[
{\rm e}^{ {\rm i}\phi}\,\left[ M_{10,00}(E,0) + M_{00,10}(E,0) \right] \Big]}
{\Im \Big[{\rm e}^{ {\rm i}\phi} \sum_{SM_S} M_{SM_S,SM_S}(E,0)\Big] 
+(p/4\pi) \sum_{SM_S} \sigma^{\rm C}_{SM_S,>}(E)} \>\>,
\label{eq:asy_t}
\end{equation}
with ${\rm exp}({\rm i}\phi)\equiv
{\rm exp}\Big[ 2{\rm i}[ \eta \, {\rm lg}\, {\rm sin}(\theta_0/2)-\sigma_0]\Big]$.
\subsection{Momentum-Space Formulation}

In order to consider the (parity-conserving) Bonn
potential~\cite{Machleidt01}, it is necessary to develop
techniques to treat the $p$$p$ scattering problem in
momentum space.  A method first proposed in Ref.~\cite{Vincent74}
and most recently applied in Ref.~\cite{Machleidt01} is used
here.  It consists in separating the potential into short-
and long-range parts $\overline{v}_{\rm S}$ and
$\overline{v}_{\rm L}$, respectively,  

\begin{equation}
\overline{v}=\overline{v}_{\rm S} + \overline{v}_{\rm L} \>\>,
\end{equation}
where

\begin{eqnarray}
\overline{v}_{\rm S}&=&\left[ v^{\rm PC}+v^{\rm PV}
+v^{\rm C}\right]\theta(R-r) \>\>, \\
\overline{v}_{\rm L}&=&v^{\rm C}\, \theta(r-R) \>\>,
\end{eqnarray}
and $\theta(x)$ is the Heaviside step function,
$\theta(x)$=1 if $x > 0$, =0 otherwise.  The radius $R$ is
chosen large enough, so that $v^{\rm PC}+v^{\rm PV}$ vanishes
for $r > R$ (in the present work, $R=20$ fm).

Since $\overline{v}_{\rm S}$ is of finite range,
standard momentum-space techniques can now be used to solve for the
$K_{\rm S}$-matrix in the $J$ channel(s):

\begin{eqnarray}
K^J_{{\rm S};\alpha^\prime,\alpha}(p^\prime;p)&=&
\overline{v}^J_{{\rm S};\alpha^\prime,\alpha}(p^\prime;p)\nonumber \\
&+&\frac{2}{\pi} \int_0^\infty {\rm d}kk^2 \sum_\beta
\overline{v}^J_{{\rm S};\alpha^\prime,\beta}(p^\prime;k)
\frac{\cal P}{p^2/(2\mu)-k^2/(2\mu)}K^J_{{\rm S};\beta,\alpha}(k;p) \>\>,
\label{eq:kma}
\end{eqnarray}
where ${\cal P}$ denotes a principal-value integration,
and the momentum-space matrix elements of the potential
$\overline{v}_{\rm S}$ are defined as in Eq.~(\ref{eq:vpe}),
but for the replacements $v\rightarrow \overline{v}_{\rm S}$
and $F_L(\eta;x)/x \rightarrow j_L(x)$.  Note that performing
the Bessel transforms of a Coulomb potential
truncated at $r=R$ poses no numerical problem.  The integral
equations~(\ref{eq:kma}) are discretized, and the resulting
systems of linear equations are solved by direct numerical
inversion.  The principal-value integration is eliminated by
a standard subtraction technique~\cite{Gloeckle83}.

The asymptotic wave functions associated with $\overline{v}_{\rm S}$
have the form:

\begin{equation}
\frac{\overline{w}^J_{{\rm S};\alpha^\prime,\alpha}(r;p)}{r}
\simeq \frac{a_\alpha}{2}
 \Big[ \delta_{\alpha,\alpha^\prime} \hat{h}^{(2)}_{L^\prime}(pr)
+\hat{h}^{(1)}_{L^\prime}(pr)
\overline{S}^J_{{\rm S};\alpha^\prime , \alpha}(p) \Big] \>\>,
\label{eq:awc}
\end{equation}
where 

\begin{equation}
\hat{h}^{(1,2)}_L(\rho)\equiv j_L(\rho) \pm {\rm i}\, n_L(\rho) \>\>,
\end{equation}
$j_L$ and $n_L$ being the regular and irregular
spherical Bessel functions, respectively, and the constants $a_\alpha$ can
only depend upon the entrance channel $\alpha$, see the Schr\"odinger
equations~(\ref{eq:cschr}).  These wave functions
should match smoothly, at $r$=$R$, those associated with the full potential
$\overline{v}_{\rm S}+\overline{v}_{\rm L}$, which behave
asymptotically as in Eq.~(\ref{eq:asy}).  Carrying out the matching
for the functions and their first
derivatives leads to a relation between the $S$-matrices
$S_{{\rm S};\alpha^\prime,\alpha}$ and $S_{\alpha^\prime,\alpha}$,
corresponding to $\overline{v}_{\rm S}$ and
$\overline{v}_{\rm S}+\overline{v}_{\rm L}$, respectively.
In terms of $K$-matrices, related to the corresponding $S$-matrices
via

\begin{equation}
S^J(p)=\left[ 1+2{\rm i} \, \mu p\, K^J(p;p) \right]^{-1}
\left[ 1-2{\rm i} \, \mu p\, K^J(p;p) \right] \>\>,
\label{eq:skma}
\end{equation}
and similarly $S_{\rm S}$ and $K_{\rm S}$,
this relation reads in matrix notation~\cite{Machleidt01a}:

\begin{eqnarray}
2 \mu p\, K^J&=&
\Bigg[ {\rm G} -\left[ {\rm J}+2 \mu p \, {\rm N} \cdot K^J_{\rm S}\right]
 \left[{\rm J}^\prime+2 \mu p \,{\rm N}^\prime \cdot K^J_{\rm S}\right]^{-1}
\cdot {\rm G}^\prime \Bigg]^{-1} \nonumber \\
&\times& \Bigg[ \left[{\rm J}+2 \mu p \, {\rm N}\cdot K^J_{\rm S}\right]
\left[ {\rm J}^\prime+2 \mu p \, {\rm N}^\prime \cdot
K^J_{\rm S}\right]^{-1}\cdot  {\rm F}^\prime-{\rm F} \Bigg] \>\>,
\end{eqnarray}
where the dependence upon $p$ is understood, and
the diagonal matrices ${\rm X}$ and ${\rm X}^\prime$ have been defined as

\begin{eqnarray}
{\rm X}_{\alpha^\prime,\alpha} &\equiv&
\delta_{\alpha^\prime,\alpha}\, X_\alpha(R) \>\>,  \\
{\rm X}^\prime_{\alpha^\prime,\alpha} &\equiv&
\delta_{\alpha^\prime,\alpha}\, \Bigg[ \frac{ {\rm d}X_\alpha(r)}
{ {\rm d}r}\Bigg]_{r=R} \>\>, 
\end{eqnarray}
with the functions 

\begin{equation}
X_\alpha(R)=j_L(pR), \>\> \frac{F_L(\eta;pR)}{pR},\>\> n_L(pR),\>\> 
{\rm and}\> \frac{G_L(\eta;pR)}{pR} 
\end{equation}
when ${\rm X}$=${\rm J}$, ${\rm F}$, ${\rm N}$, and ${\rm G}$, respectively.
Once the $K$-matrices in the various channels have been determined,
the corresponding $S$-matrices are obtained from Eq.~(\ref{eq:skma}), from
which the amplitude $M_{S^\prime M^\prime_S,SM_S}(E,\theta)$~(\ref{eq:am})
is constructed.
\subsection{Matrix elements of $v^{\rm PV}$ in channel $J$}

To evaluate the radial functions $v^{J,\, {\rm PV}}_{\alpha^\prime,\alpha}(r)$
of the PV potential in Eq.~(\ref{eq:vpe})--those associated with
the PC potential are well known--one needs the matrix
elements of $({\bbox \sigma}_1 \times{\bbox \sigma}_2)\cdot \hat{\bf r}$
and $({\bbox \sigma}_1- {\bbox \sigma}_2) \cdot {\bf p}$ between spin-angle
functions.  Using the notation

\begin{equation}
\langle J;L^\prime, S^\prime \mid O \mid J; L,S\rangle \equiv
\int{\rm d}\Omega\, {\cal Y}_{L^\prime S^\prime J}^{M_J \dagger}
\, O({\bf r}) \, {\cal Y}_{LSJ}^{M_J} \>\>,
\end{equation}
and writing 

\begin{equation}
({\bbox \sigma}_1- {\bbox \sigma}_2) \cdot {\bf p}=-{\rm i} \,
({\bbox \sigma}_1- {\bbox \sigma}_2) \cdot \left[\hat{\bf r} 
\frac{\vec{\partial}}{\partial r} + \frac{1}{r} \frac{\vec{\partial}}
{\partial {\bbox \Omega}} \right] \>\>,
\end{equation}
where the $\vec{\partial}$ symbol indicates that the partial
derivatives must act to the right, one finds that the
non-vanishing matrix elements are:

\begin{eqnarray}
\langle J;J,0 \mid 
({\bbox \sigma}_1 \times{\bbox \sigma}_2)\cdot \hat{\bf r} \mid J;J\mp 1,1\rangle
&=&\pm {\rm i}\, \sqrt{\frac{2\, J+1\mp 1}{J +1/2}} \>\>, \label{eq:vpv1}\\
\langle J;J,0 \mid 
({\bbox \sigma}_1 - {\bbox \sigma}_2)\cdot \hat{\bf r} \mid J;J\mp 1,1\rangle
&=&\pm \sqrt{\frac{2\, J+1\mp 1}{J +1/2}} \>\>,
\label{eq:vpv2}
\end{eqnarray}

\begin{eqnarray}
\langle J;J,0 \mid
({\bbox \sigma}_1 - {\bbox \sigma}_2)\cdot \frac{\vec{\partial}}
{\partial {\bbox \Omega}} \mid J;J\mp 1,1\rangle &=&
-\frac{2\, J+1 \mp 3}{2} \sqrt{\frac{2\, J+1\mp 1}{J +1/2}} \>\>, \label{eq:vpv3} \\
\langle J;J\mp 1,1 \mid
({\bbox \sigma}_1 - {\bbox \sigma}_2)\cdot \frac{\vec{\partial}}
{\partial {\bbox \Omega}} \mid J;J,0\rangle &=&
\frac{2\, J+1 \pm 1}{2} \sqrt{\frac{2\, J+1\mp 1}{J +1/2}} \>\>.
\label{eq:vpv4}
\end{eqnarray}
Note that the operators in Eqs.~(\ref{eq:vpv1})--(\ref{eq:vpv2}) are Hermitian,
while those in Eqs.~(\ref{eq:vpv3})--(\ref{eq:vpv4}) are not.
The complete Hamiltonian is, of course, Hermitian.

\section{Results and Discussion}
\label{sec:res}

In this section we present results for the longitudinal asymmetry in
the lab energy range 0--350 MeV.  The calculations use any of the
modern strong-interaction potentials, either Argonne $v_{18}$ (AV18)~\cite{Wiringa95},
or CD-Bonn (BONN)~\cite{Machleidt01}, or Nijmegen-I (NIJ-I)~\cite{Stoks94}, 
in combination with the DDH weak-interaction potential parameterized in terms of
$\rho$- and $\omega$-meson exchanges~\cite{Desplanques80}.  The values
for the $\rho$- and $\omega$-meson coupling constants and cutoff parameters
are listed in Table~\ref{tb:gs}.  The strong-interaction coupling
constants and cutoff parameters are taken from the BONN potential, while
the weak-interaction coupling constants $h^{pp}_\rho$ and
$h^{pp}_\omega$ have been determined by an AV18-based fit to the
observed asymmetry.  In Table~\ref{tb:gs} we also list
the $h^{pp}_\rho$ and $h^{pp}_\omega$ values corresponding to the \lq\lq best\rq\rq
estimates for the $h_{\rho_i}$ and $h_{\omega_i}$ suggested in Ref.~\cite{Desplanques80},
column labeled DDH-orig. 

The data points for the longitudinal asymmetry at 13.6 MeV, 45 MeV,
and 221 MeV are those reported in Ref.~\cite{Berdoz01,vanOers01}, and their values
are $(-0.97\pm 0.20)\times 10^{-7}$,  $(-1.53\pm 0.21)\times 10^{-7}$, and
$(+0.84\pm 0.34)\times 10^{-7}$, respectively.  The first point at 13.6
MeV has been obtained~\cite{vanOers01} by taking the weighted mean--and accounting
for the square-root energy dependence--of the latest result from the the Bonn
experiment at 13.6 MeV, as reported by Eversheim (Ref.~[14]
in Ref.~\cite{Berdoz01}), and the 15 MeV result from Ref.~\cite{Nagle79}.
The point at 45 MeV has also been obtained~\cite{vanOers01} by combining
results from measurements at 45 MeV~\cite{Kistryn87}, 46 MeV, and 47 MeV
(these last two both from Ref.~\cite{vanOers01}).  The last point at
221 MeV is that reported in Ref.~\cite{Berdoz01}.  Finally, the errors
include both statistical and systematic errors added in quadrature.

\begin{figure}[bth]
\let\picnaturalsize=N
\def\picsize{4in}
\def\picfilenamea{asy_pp.eps}
\ifx\nopictures Y\else{\ifx\epsfloaded Y\else\input epsf \fi
\let\epsfloaded=Y
\centerline{
\ifx\picnaturalsize N\epsfxsize \picsize\fi \epsfbox{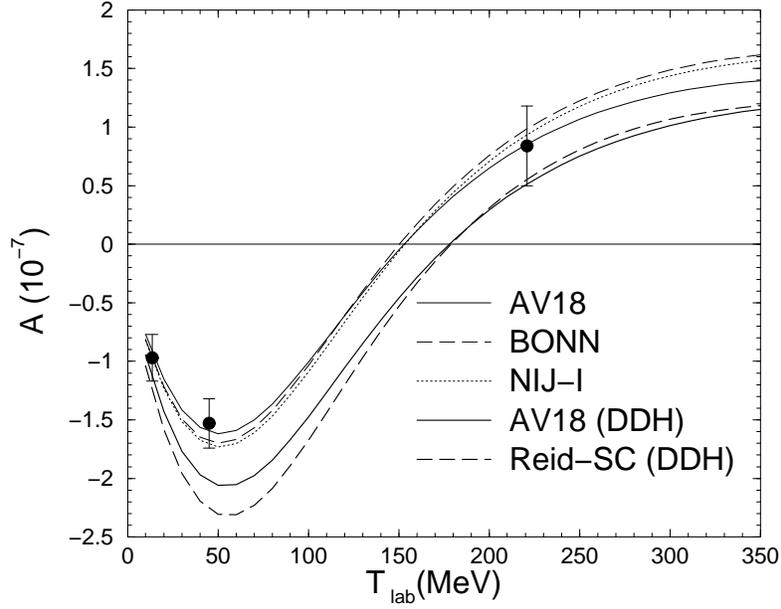}
 }}\fi
\caption{Total nuclear asymmetries obtained with various combinations
of strong- and weak-interaction potentials are compared to data, see text.}
\label{fig:asy}
\end{figure}

The total longitudinal asymmetry, shown in Fig.~\ref{fig:asy}
for a number of combinations of strong- and weak-interaction
potentials, is defined as 

\begin{equation}
A(E)= \frac{\Im \Big[ M_{10,00}(E,0) + M_{00,10}(E,0)\Big]}
{\Im \Big[\sum_{SM_S} M_{SM_S,SM_S}(E,0)\Big]} \>\>,
\end{equation}
where the amplitudes $M_{S^\prime M_S^\prime ,SM_S}(E,\theta)$
are those given in Eq.~(\ref{eq:am}).  The expression above
for $A(E)$ ignores the contribution of the Coulomb amplitude,
Eq.~(\ref{eq:amc}), divergent in the limit $\theta$=0, and for this reason
$A(E)$ will be referred to as the \lq\lq nuclear\rq\rq asymmetry. 
Of course, one should note that Coulomb potential effects
enter into $A(E)$ explicitly through the Coulomb phase shifts, present
in the partial-wave expansion for $M_{S^\prime M_S^\prime,SM_S}(E,\theta)$,
and implicitly through the wave functions, from which the $S$-matrix
elements are calculated.  The effect of including explicitly
the amplitude induced by the Coulomb potential is discussed below. 

The calculated nuclear asymmetries in Fig.~\ref{fig:asy} were  
obtained by retaining in the partial-wave expansion for
$M_{S^\prime M_S^\prime ,SM_S}(E,\theta)$ all channels with
$J$ up to $J_{\rm max}=8$.  The curves labeled AV18, BONN, and
NIJ-I all use the DDH potential with the coupling constants $h^{pp}_\rho$
and $h^{pp}_\omega$ determined by a rough fit to data (the AV18 is used
in the fitting procedure).  There is very little sensitivity
to the input strong-interaction potential, certainly
much less than one would infer from Fig.~1 of Ref.~\cite{Driscoll89}.
This is undoubtedly a consequence of the more extended $p$$p$ and
$p$$n$ scattering data-base to which present potentials are fitted, as
well as of the much higher accuracy achieved in these fits.
An analysis of the extracted $h^{pp}_\rho$ and $h^{pp}_\omega$
coupling constants and their errors is presented later in this section.
 
We also show in Fig.~\ref{fig:asy} the AV18 results which
correspond to a DDH potential using the \lq\lq best\rq\rq estimates for the 
$h^{pp}_\rho$ and $h^{pp}_\omega$ coupling constants~\cite{Desplanques80}
(values in column DDH-orig in Table~\ref{tb:gs}), with the remaining $\rho$-
and $\omega$-meson strong-interaction coupling constants and cutoff
parameters as given in Table~\ref{tb:gs}.  A number of comments are now
in order.  The data point at 221 MeV essentially determines the value
of $h^{pp}_\rho$.  As pointed out by Simonius~\cite{Simonius88} (see also
below), the dominant contributions to the total asymmetry in the energy
range under consideration here are those associated with the $^1$S$_0$-$^3$P$_0$
and $^3$P$_2$-$^1$D$_2$ partial waves.  At energies close to 225 MeV
the $^1$S$_0$-$^3$P$_0$ contribution, which can easily be shown to be
proportional to ${\rm cos}\left[\delta(E;^1{\rm S}_0) + \sigma_1(E) +\sigma_0(E) \right] 
-{\rm cos}\left[\delta(E;^3{\rm P}_0) + \sigma_1(E) +\sigma_0(E) \right]$ using
Eq.~(\ref{eq:am}) (here $\delta(E;^1{\rm S}_0)$ and $\delta(E;^3{\rm P}_0)$
are the strong-interaction phases), vanishes.  As a result, the total asymmetry
in this energy region is almost entirely due to the $^3$P$_2$-$^1$D$_2$ contribution,
which is known~\cite{Simonius88} to be approximately proportional
to the following combination of coupling constants,
$h^{pp}_\rho g_\rho \kappa_\rho+ h^{pp}_\omega g_\omega \kappa_\omega$.
In the BONN model, the $\omega$-meson tensor coupling constant is taken to be zero,
and hence the data point at 221 MeV fixes $h^{pp}_\rho$ (for given
$g_\rho$, $\kappa_\rho$, and $\Lambda_\rho$).  This is the reason for the
$\simeq 44$ \% increase (in magnitude) of $h^{pp}_\rho$ with respect
to the DDH \lq\lq best\rq\rq estimate.

Below 50 MeV, however, the calculated total asymmetry
is dominated by the $^1$S$_0$-$^3$P$_0$ contribution,
approximately proportional to~\cite{Simonius88}
$h^{pp}_\rho g_\rho ( 2 +\kappa_\rho )
+h^{pp}_\omega g_\omega ( 2 +\kappa_\omega )$.  The increase
in magnitude of $h^{pp}_\rho$ required to fit the point at 221 MeV,
now leads to a total asymmetry $\mid\!\! A(E)\!\!\mid$ below 50 MeV, which is too large
when compared to experiment.  Thus, in order to reproduce
the 13.6 MeV and 45 MeV data points,
the overall strength of the coupling constant combination above needs
to be reduced significantly.  Since $g_\rho ( 2 +\kappa_\rho )$ and
$g_\omega ( 2 +\kappa_\omega )= 2\, g_\omega$ have the same sign, this
requires making the sign of $h^{pp}_\omega$ opposite to that of $h^{pp}_\rho$.

It is worth pointing out, though, that the changes in value
for $h^{pp}_\rho$ and $h^{pp}_\omega$ advocated here are still compatible
with the \lq\lq reasonable\rq\rq ranges for the $h^{pp}_{\rho_i}$ and
$h^{pp}_{\omega_i}$, determined in Ref.~\cite{Desplanques80}.

Finally, we show in Fig.~\ref{fig:asy} the total nuclear asymmetry
obtained in a calculation based on the old Reid soft-core
potential~\cite{Reid68,Wiringa01} and a DDH potential using
the following coupling-constant and cutoff values: $g^2_\rho/4\pi = 0.95$,
$g^2_\omega/4\pi = 20$, $\kappa_\rho=6.1$, $\kappa_\omega=0$,
$\Lambda_\rho=1.3$ GeV/c, $\Lambda_\omega=1.5$ GeV/c (these are all
from the old $r$-space version of the Bonn potential~\cite{Machleidt87}),
and the \lq\lq best\rq\rq estimates for $h^{pp}_\rho$ and $h^{pp}_\omega$.
These model interactions are essentially identical to those employed by Driscoll and Miller
in Ref.~\cite{Driscoll89}.  Indeed, our calculated total asymmetry
is close to that obtained by these authors.  It should be stressed that
in Ref.~\cite{Driscoll89} the strong-interaction phases and mixing angles
were taken from Arndt's analysis of nucleon-nucleon
scattering data~\cite{Arndt87} rather than calculated from
the Reid soft-core potential, as done here.  This is presumably
the origin of the remaining small differences between their results
and ours.

\begin{figure}[bth]
\let\picnaturalsize=N
\def\picsize{4in}
\def\picfilenamea{asy_pp_j.eps}
\ifx\nopictures Y\else{\ifx\epsfloaded Y\else\input epsf \fi
\let\epsfloaded=Y
\centerline{
\ifx\picnaturalsize N\epsfxsize \picsize\fi \epsfbox{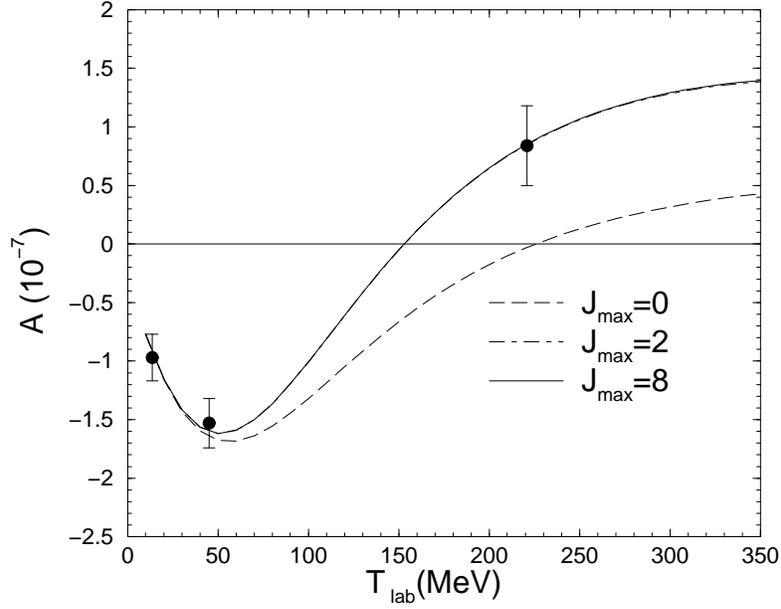}
 }}\fi
\caption{Contributions to the total nuclear asymmetry
obtained by including only the $J$=0 channel, and by adding the
$J$=2 channels, and finally all even $J$-channels up to $J_{\rm max}$=8.
The AV18- and DDH-adj potential combination is used, thin solid line
in Fig.~\protect{\ref{fig:asy}}.}
\label{fig:asy_j}
\end{figure}

Figure~\ref{fig:asy_j} shows the total nuclear asymmetries
obtained by including only the $J$=0 channel ($^1$S$_0$-$^3$P$_0$)
and, in addition, the $J$=2 channels ($^3$P$_2$-$^1$D$_2$ 
and $^3$F$_2$-$^1$D$_2$), and finally all (even) $J$-channels 
up to $J_{\rm max}$=8.  We re-emphasize that in the energy range
0--350 MeV the asymmetry is dominated by the $J=$0 and 2 contributions
(among the latter, specifically those from the
$^3$P$_2$-$^1$D$_2$ partial waves).

\begin{figure}[bth]
\let\picnaturalsize=N
\def\picsize{4in}
\def\picfilenamea{asy_pp_lnew.eps}
\ifx\nopictures Y\else{\ifx\epsfloaded Y\else\input epsf \fi
\let\epsfloaded=Y
\centerline{
\ifx\picnaturalsize N\epsfxsize \picsize\fi \epsfbox{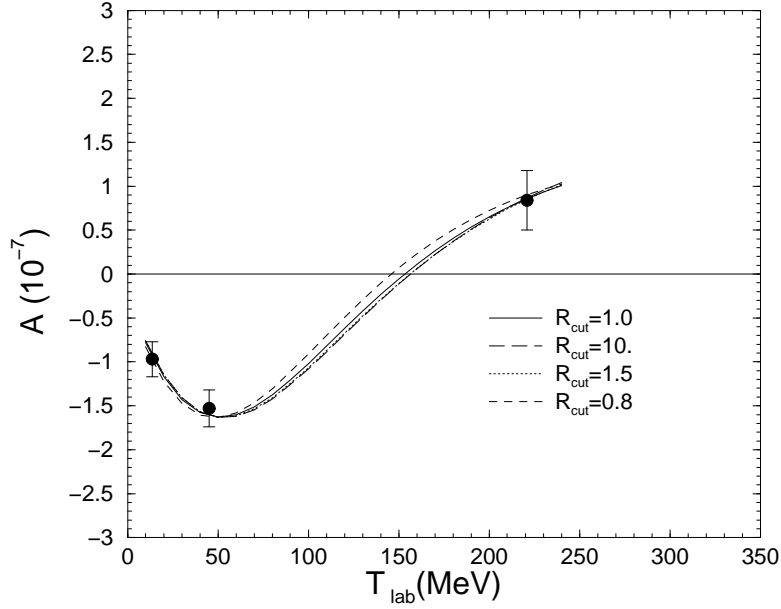}
 }}\fi
\caption{Sensitivity of the total nuclear asymmetry to modifications of
the cutoff-parameters $\Lambda_\rho$ and $\Lambda_\omega$ in the DDH potential.
Both cut-offs are multiplied by $R_{\rm cut}$, see text.  For each case
the couplings are then adjusted to approximately reproduce the
results obtained with the AV18 and DDH-adj potential combination (corresponding
to $R_{\rm cut}$=1) in Fig.~\protect{\ref{fig:asy}}.}
\label{fig:asy_l}
\end{figure}

Figure~\ref{fig:asy_l} illustrates the sensitivity of the total
nuclear asymmetry to modifications of the $\Lambda_{\rho}$ and $\Lambda_\omega$
cutoff parameters in the DDH potential.  
Both cot-offs are multiplied by $R_{\rm cut}$, in each case
$h^{pp}_\rho$ and $h^{pp}_\omega$ are then readjusted to approximately reproduce
the AV18+DDH-adj combination.  In the near point-like
limit ($R_{\rm cut}$=10), the asymmetry increases in magnitude
by roughly a factor of 2 prior to adjustment.
The resulting couplings used for this case are
$h^{pp}_\rho$=--14.33 and $h^{pp}_\omega$=+3.95.
Results are also shown for $R_{\rm cut}$=1.5 and 0.8, the
latter is an extreme case where the cutoff parameters are 
near the meson masses used to determine the ranges.
Nevertheless, the energy dependence of
the asymmetry is in all cases very similar.
The couplings used to generate the two other curves are:
for $R_{\rm cut}$=1.5, $h^{pp}_\rho$=--15.32 and $h^{pp}_\omega$=+3.92;
and for $R_{\rm cut}$=0.8, $h^{pp}_\rho$=--106.7 and $h^{pp}_\omega$=+14.63.

\begin{figure}[bth]
\let\picnaturalsize=N
\def\picsize{4in}
\def\picfilenamea{asy_coul.eps}
\ifx\nopictures Y\else{\ifx\epsfloaded Y\else\input epsf \fi
\let\epsfloaded=Y
\centerline{
\ifx\picnaturalsize N\epsfxsize \picsize\fi \epsfbox{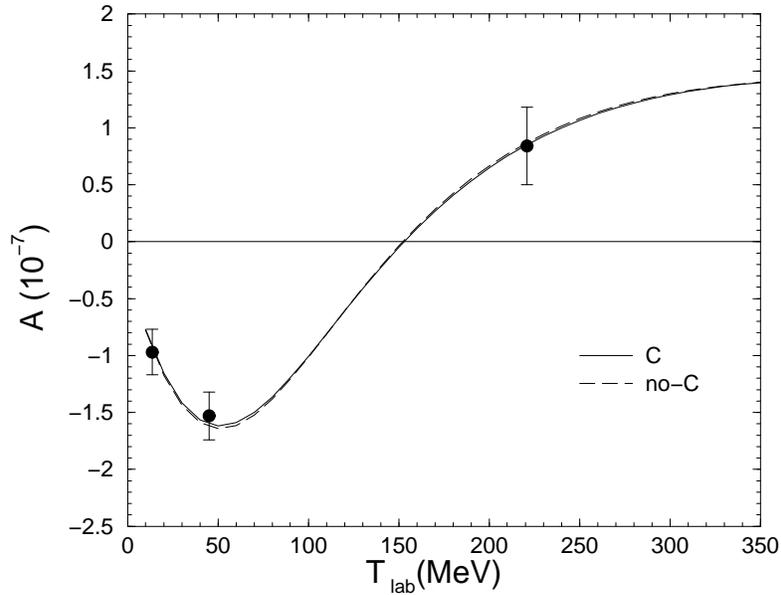}
 }}\fi
\caption{The total nuclear asymmetry, see text, is compared to the total
asymmetry obtained by ignoring the Coulomb potential.
The AV18 and DDH-adj potential combination is used, thin solid line
in Fig.~\protect{\ref{fig:asy}}.}
\label{fig:asy_coul}
\end{figure}

\begin{figure}[bth]
\let\picnaturalsize=N
\def\picsize{4in}
\def\picfilenamea{asy_pp_ang.eps}
\ifx\nopictures Y\else{\ifx\epsfloaded Y\else\input epsf \fi
\let\epsfloaded=Y
\centerline{
\ifx\picnaturalsize N\epsfxsize \picsize\fi \epsfbox{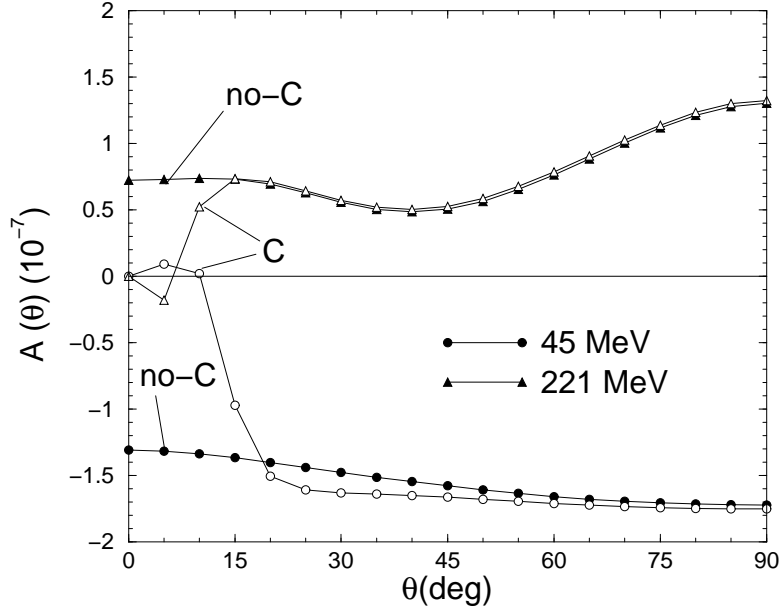}
 }}\fi
\caption{Angular distributions obtained with (curve labeled C)
and without (curve labeled no-C) inclusion of the Coulomb
amplitude at 45 MeV and 221 MeV, see text.
The AV18 and DDH-adj potential combination is used, thin solid line
in Fig.~\protect{\ref{fig:asy}}.}
\label{fig:asy_ang}
\end{figure}

\begin{figure}[bth]
\let\picnaturalsize=N
\def\picsize{4in}
\def\picfilenamea{asy_pp_tra.eps}
\ifx\nopictures Y\else{\ifx\epsfloaded Y\else\input epsf \fi
\let\epsfloaded=Y
\centerline{
\ifx\picnaturalsize N\epsfxsize \picsize\fi \epsfbox{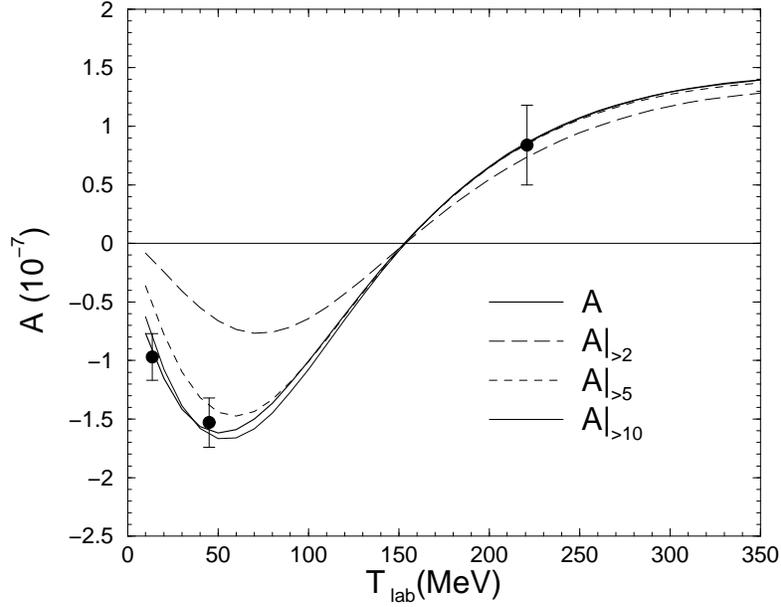}
 }}\fi
\caption{The total nuclear asymmetry (thick solid line) is compared
with the total asymmetries measured in transmission experiments with critical
angles $\theta_0$=2$^\circ$, 5$^\circ$, and 10$^\circ$ (curves
labeled $A\mid_{>2}$, $A\mid_{>5}$, and $A\mid_{>10}$, respectively). 
The AV18 and DDH-adj potential combination is used, thin solid line
in Fig.~\protect{\ref{fig:asy}}.}
\label{fig:asy_tra}
\end{figure}

Figures~\ref{fig:asy_coul}--\ref{fig:asy_tra} illustrate the effects
of Coulomb contributions on the longitudinal asymmetry.  Figure~\ref{fig:asy_coul}
compares the total nuclear asymmetry defined above (curve labeled \lq\lq C\rq\rq)
with the total asymmetry obtained by ignoring the Coulomb potential
altogether (curve labeled \lq\lq no-C\rq\rq).  As already pointed out
in Ref.~\cite{Driscoll89}, Coulomb contributions to these (non-physical)
quantities are rather small.

Figure~\ref{fig:asy_ang} compares the angular distribution
of the (physical) longitudinal asymmetry obtained from the amplitudes
$\overline{M}=M+M^{\rm C}$, see Eq.~(\ref{eq:angc}), with that 
calculated by replacing $\overline{M} \rightarrow M$ in Eq.(\ref{eq:angc}),
namely ignoring the contribution of the Coulomb amplitude $M^{\rm C}$. 
The latter dominates at small scattering angles, and leads to the
peculiar small angle behavior of the angular
distribution shown in Fig.~\ref{fig:asy_ang}, namely its changing of sign
at small $\theta$, and its vanishing at $\theta$=0
(also observed in Ref.~\cite{Driscoll89}).  It is interesting
to note that the angular distribution at 230 MeV obtained by
the authors of Ref.~\cite{Driscoll89} is significantly different
from that shown in Fig.~\ref{fig:asy_ang} at 221 MeV.

\begin{figure}[bth]
\let\picnaturalsize=N
\def\picsize{4in}
\def\picfilenamea{asy_pp_b.eps}
\ifx\nopictures Y\else{\ifx\epsfloaded Y\else\input epsf \fi
\let\epsfloaded=Y
\centerline{
\ifx\picnaturalsize N\epsfxsize \picsize\fi \epsfbox{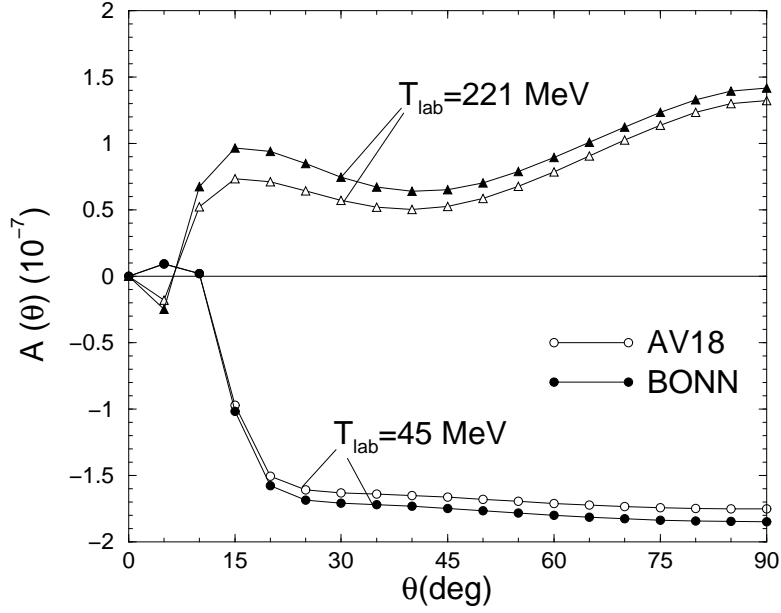}
 }}\fi
\caption{Angular distributions of the longitudinal asymmetry
obtained with either the AV18 or BONN potential in combination with
the DDH-adj potential.}
\label{fig:asy_bonn}
\end{figure}

Figure~\ref{fig:asy_tra} illustrates the effects of Coulomb contributions
on the total longitudinal asymmetry measured in
transmission experiments, Eq.~(\ref{eq:asy_t}), for
various choices of the critical angle $\theta_0$ ($\theta_0$=2$^\circ$, 5$^\circ$,
and 10$^\circ$), see Eq.~(\ref{eq:s_tra}).  Coulomb contributions are
substantial, particularly for small $\theta_0$ and energies below 100 MeV. 
We also demonstrate, in Fig.~\ref{fig:asy_bonn}, that the
angular distributions of the longitudinal asymmetry are
only weakly affected by different input (strong-interaction)
potentials.

Finally, we present an analysis of the extracted coupling constants and
their experimental errors.  This analysis employs the AV18 model
and the BONN-derived strong interaction couplings and cut-offs 
(Table~\ref{tb:gs}) in the DDH potential.  Given the weak sensitivity
described earlier, only small changes should be expected with the
use of other recent strong interaction potentials.
The experimental data at low energies have been combined into the two
data points shown in Fig.~\ref{fig:asy} at 13.6 and 45 MeV.
The asymmetries are --0.97 $\pm$ 0.2 and --1.53 $\pm$ 0.21, respectively,
combining statistical and systematic errors.
We also include the recent TRIUMF result of +0.84 $\pm$ 0.34
at 221 MeV (all in units of $10^{-7}$).

\begin{figure}[bth]
\let\picnaturalsize=N
\def\picsize{4in}
\def\picfilenamea{contour-bw.eps}
\ifx\nopictures Y\else{\ifx\epsfloaded Y\else\input epsf \fi
\let\epsfloaded=Y
\centerline{
\ifx\picnaturalsize N\epsfxsize \picsize\fi \epsfbox{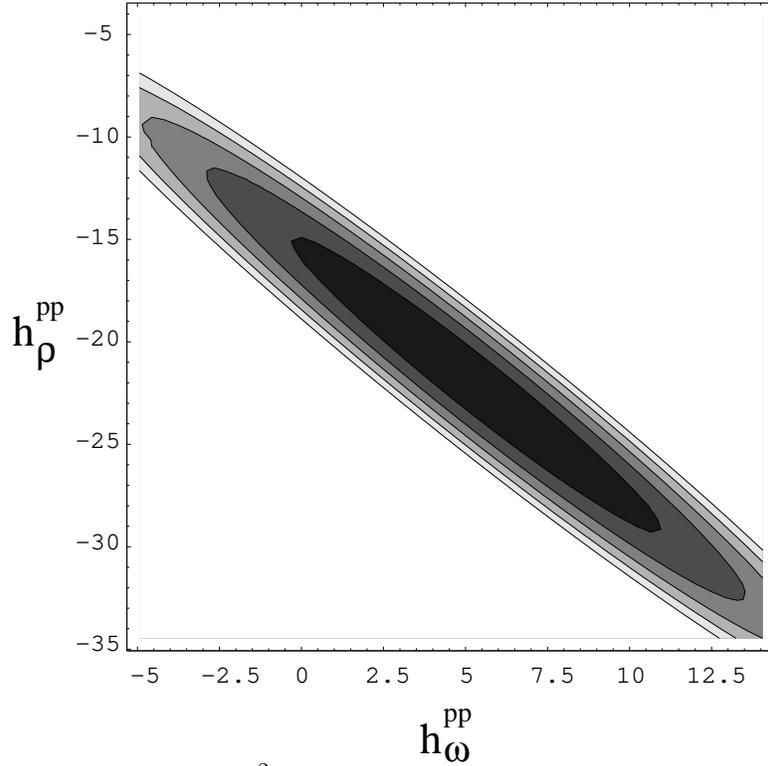}
 }}\fi
\caption{Curves of constant total $\chi^2$ obtained by
analyzing the experimental $pp$ data with the AV18 model,
and $\rho$- and $\omega$-meson strong-interaction couplings in the DDH potential
from the Bonn-2000 model.  The curves indicate surfaces
of total $\chi^2$ = 1,2,3,4, and 5 for various values of
$h^{pp}_\rho$ and $h^{pp}_\omega$.}
\label{fig:errcontour}
\end{figure}

Figure~\ref{fig:errcontour} shows contours of constant total
$\chi^2$ at levels of 1 through 5 versus the coupling
constants $h^{pp}_\rho$ and $h^{pp}_\omega$.
As is apparent in the
figure, there is a rather narrow band of acceptable values for
$h^{pp}_\rho$ and $h^{pp}_\omega$ at total $\chi^2 = 1$ for the
3 experimental data points.  At this level, $h^{pp}_\rho$ can range
from approximately --14 to --28, with a simultaneous
(and strongly-correlated) variation in
$h^{pp}_\omega$ from rougly --2 to +10, all in units of $10^{-7}$.

\section{Conclusions}

We have performed an analysis of the $pp$ parity-violating (PV)
longitudinal asymmetry using combinations of modern-day strong
interaction potentials and the DDH PV potential.  The
new experimental results from TRIUMF at 221 MeV, in combination
with previous results at lower energy, provide a strong constraint
on allowable linear combinations of $\rho$ and $\omega$ 
PV coupling constants.  Combining the statistical and
systematic errors in quadrature, $h^{pp}_\rho$ is
constrained by present data to approximately 35\%,
at the level of one standard deviation.

The prime motivation for the present work is to initiate a 
systematic and consistent study of many PV 
observables in the few-nucleon sector, where accurate
microscopic calculations are feasible.
Recent measurements of nuclear anapole moments
in atomic PV experiments are difficult to reconcile with
earlier PV experiments in light nuclei
using the simple DDH-orig model~\cite{Haxton01}.
Up to now, this approach has been extremely limited
by the available data.

The present experimental data in the few-nucleon sector
remain rather sparse, with primary constraints coming from
the $p$$p$ longitudinal asymmetry
analyzed here and measurements of the PV asymmetry in
$p-\alpha$ elastic scattering~\cite{Lang86}.
A variety of new results are expected in the next few
years, however, including $^1H(\vec{n},\gamma)^2H$~\cite{Snow00},
the neutron spin rotation in Helium~\cite{Heckelpc}, and electron
scattering measurements at Bates and JLAB.
The combination of these diverse experiments should finally yield
a coherent picture of the NN PV interaction
at the hadronic scale.

\label{sec:cons}

\section*{Acknowledgments}
We wish to thank R.\ Machleidt for providing us with computer codes
generating the latest version of the Bonn potential, and
for illuminating correspondence in regard to the treatment
of the Coulomb potential in momentum space.
The work of J.C. and B.F.G. was supported by the 
U.S. Department of Energy under contract W-7405-ENG-36, and
the work of R.S. was supported by DOE contract DE-AC05-84ER40150
under which the Southeastern Universities Research Association (SURA)
operates the Thomas Jefferson National Accelerator Facility.
Finally, some of the calculations were made possible by grants
of computing time from the National Energy Research Supercomputer
Center.
\begin{table}
\caption{Values used for the strong- and weak-interaction
coupling constants of the $\rho$- and $\omega$-meson to the nucleon, see text.}
\begin{tabular}{cccccc}
 & $g^2_\alpha/4\pi$  & $\kappa_\alpha$  & $10^7 h_\alpha^{pp}$ (DDH-adj)  & $10^7 h_\alpha^{pp}$ (DDH-orig) & $\Lambda_\alpha$ (GeV/c)  \\
\tableline
$\rho$   & 0.84 & 6.1 & --22.3 & --15.5  & 1.31 \\ 
$\omega$ & 20. & 0.  & +5.17 & --3.04   & 1.50 
\end{tabular}
\label{tb:gs}
\end{table}
\begin{table}
\caption{Labeling of channels.}
\begin{tabular}{cccc}
         & \multicolumn{3}{c} {$\alpha$} \\
$J$      & 1  & 2 & 3  \\
\tableline
0        & $^1$S$_0$ & $^3$P$_0$ &           \\
1        & $^3$P$_1$ &           &           \\
2        & $^3$P$_2$ & $^3$F$_2$ & $^1$D$_2$ \\
3        & $^3$F$_3$ &           &           \\
4        & $^3$F$_4$ & $^3$H$_4$ & $^1$G$_4$ \\
$\dots$  & $\dots$   & $\dots$   & $\dots$
\end{tabular}
\label{tb:chan}
\end{table}
\begin{table}
\caption{Classification of channel mixings for $J$ even: PC or PV
if induced by $v^{\rm PC}$ or $v^{\rm PV}$, respectively.}
\begin{tabular}{cccc}
         & \multicolumn{3}{c} {coupling} \\
$J$      & 12  & 13 & 23  \\
\tableline
0         & PV &    &     \\
2         & PC & PV & PV  \\
4         & PC & PV & PV  \\
$\dots$   & PC & PV & PV 
\end{tabular}
\label{tb:mixing}
\end{table}
\end{document}